\documentstyle[aps,preprint]{revtex}

\author{Diego F. Torres\thanks{e-mail: dtorres@venus.fisica.unlp.edu.ar}}
\address{Departamento de F\'{\i}sica, Universidad Nacional de La Plata \\ 
C.C. 67, 1900, La Plata, Buenos Aires, Argentina}
\title{Slow Roll Inflation in Non-Minimally Coupled Theories: Hyperextended
Gravity Approach}

\begin{document}

\maketitle

\begin{abstract}
The slow roll approximation is studied for cosmological models in
Hyperextended Scalar-Tensor Theories of Gravity.
A procedure to obtain slow roll solutions in non-minimally
coupled gravity is outlined and some examples are provided.
An integral condition over the functional form of the
non-minimal coupling is imposed in order to obtain intermediate
inflationary behavior.

\end{abstract}
\newpage

\section{Introduction}

In this work we analize the slow roll approximation in cosmological
models of non-minimally coupled theories of gravity (NMC) starting from the 
complete two free functions lagrangian density of Hyperextended Scalar-Tensor
Theories (HSTG) \cite{Torres}. These HSTG are different from both, 
generalized Brans-Dicke (BD) cases and and NMC cases, 
when there are non-invertible
functions of the field involved in the lagrangian. This give rise to the 
representation of a singularity in the field transformation which allows
the passage from a NMC lagrangian to a BD one 
\cite{LiddleWands}. HSTG 
may reduce itself either
to BD or to NMC whenever convenient. In past work, exact 
cosmological solutions were obtained for these models in the case of vacuum, 
radiation and stiff matter universes \cite {Torres}. 
Concerning inflation, exact and approximate solutions were found
for particular scalar-tensor theories \cite{Solutions}. The slow roll
approximation \cite {SR} were analized by Garcia Bellido, Linde and Linde
in the case of BD gravity \cite{Garcia} and later by Barrow for generalized BD
gravity with power law couplings \cite{BarrowSR}.
Here, we are going to present 
the slow roll approximation in the context of HSTG. We work
out, following the method used by Barrow, examples for non-minimally 
coupled gravity.  We impose an
integral vinculum over the form of the non-minimal coupling in order to have
intermediate inflation behavior.

\section{Hyperextended Scalar-Tensor Theories and the Slow Roll Approximation}

The hyperextended scalar-tensor field equations are derived from the action:

\begin{equation}
\label{7}S=\int d\Omega \ \sqrt{-g}\ \left[ 16\pi L_M+\frac{K(\phi )}2\phi
_{,\mu }\,\phi ^{,\mu }+G(\phi )^{-1}\,R\right]
\end{equation}
From now on, we shall call $K(\phi )=\frac{-2\omega (\phi )}\phi $ to
facilitate the comparison with the BD cases.
This must be understood only as a change in the names of the functions. 
Taking variational derivatives
from (1) with respect to the dynamical variables $g^{\mu \nu }$ and $\phi $
we get the field equations.
We are interested in an isotropical
and homogeneous universe. We shall have a metric defined by a Friedmann-Robertson-walker one
with zero curvature and a stress-energy tensor for matter given by the deduced from
a lagrangian density of another scalar field $\sigma$ with self interaction potential
$V(\sigma)$. Finally, these equations become:

\begin{equation}
H^2-H \frac1G\frac{dG}{d\phi }\dot \phi -\frac \omega 6\frac{\dot \phi ^2}\phi G
=\frac{8\pi }3 G\left[\frac12 \dot \sigma^2 + V(\sigma)\right]
\end{equation}

\begin{equation}
\dot \phi ^2 \Gamma
+\Box \phi \left[ \frac{2\omega }\phi +\frac
3{G^3}\left( \frac{dG}{d\phi }\right) ^2\right]=
-\frac 1G\frac{dG}{d\phi }8\pi \left[4 V(\sigma)-\dot\sigma^2\right]
\end{equation}

\begin{equation}
\label{4}
\ddot\sigma +3 H \dot\sigma + V^{\prime}(\sigma)=0
\end{equation}
where we have defined
\begin{equation}
\Gamma=\left[ \frac 1\phi \frac{d\omega }{d\phi }-\frac
\omega {\phi ^2}-\frac 1G\frac{dG}{d\phi }\frac \omega \phi -\frac
6{G^4}\left( \frac{dG}{d\phi }\right) ^3+\frac 3{G^3}\frac{dG}{d\phi }\frac{d^2G}{d\phi ^2}\right] 
\end{equation}
and $H$ as usual. These equations reduce to those of GR when $\phi$
equals to a constant and to those of BD when $G$ equals $1/\phi$.

In order to examine the forms of inflation that can arise in these kind of models, 
we are going to assume the slow roll approximation given by:

\begin{equation}
\ddot \sigma \ll H \dot \sigma
\end{equation}

\begin{equation}
\frac 12 \dot\sigma^2 \ll V(\sigma)
\end{equation}

\begin{equation}
\label{A1}
\ddot\phi \ll H \dot\phi \ll H^2 \phi
\end{equation}
The last condition requires a scalar field which evolves slowly with respect to the
expansion of the universe and was studied and applied  in the context of both,
BD gravity \cite {Garcia} and generalized BD gravity \cite{BarrowSR}.
Condition (\ref{A1}) may be used to fix a similar one, suitable to be 
applied in the more general case. It is possible to see that, for any
function $f$ 
positive defined and with a convergent Taylor serie,
condition (\ref{A1}) may be transformed to

\begin{equation}
\label{A2}
\ddot f(\phi)\ll H \dot f(\phi) \ll H^2 f(\phi)
\end{equation}
in any case in which $\sim \exists n / n \in {\bf N} \wedge n \dot \phi \simeq H \phi$.
In particular, the choice $f=G^{-1}$ makes possible to fix useful simplifications 
on the field equations (2-4). 
Finally, the slow roll field equations become:

\begin{equation}
\label{SR1}
3H\dot\sigma \simeq - V^{\prime}(\sigma)
\end{equation}

\begin{equation}
\label{SR2}
3H^2\simeq\frac \omega 2\frac{\dot \phi ^2}\phi G
+\frac{8\pi }3 G V(\sigma)
\end{equation}

\begin{equation}
\label{SR3}
\dot \phi ^2 \Gamma+
 3 H\dot\phi \left[ \frac{2\omega }\phi +\frac
3{G^3}\left( \frac{dG}{d\phi }\right) ^2\right]\simeq
-\frac 1G\frac{dG}{d\phi }32\pi V(\sigma)
\end{equation}
The leading order terms referred in the slow roll equations also reduce 
to the BD ones when $G=1/\phi$ as expected (see reference \cite{BarrowSR}
for comparison). From here, and in order to proceed further, it is necessary
to define both, $\omega(\phi)$ and $G(\phi)$ together with the form of the potential
$V(\sigma)$ for the inflaton.

In \cite{BarrowSR}, a complete fixing of the problem was obtained by 
defining the functional form of the coupling, in that case a power law.
We propose now, using the freedom given by having two generic functions
instead of one, to sketch how general analytical solutions for the slow roll
equations for non-minimally coupled theories may be obtained.
From now on, we are going to use equalities even when we only have 
approximate expressions.

\section{Slow Roll Solutions of Non-minimally Coupled Gravity}

The NMC theories of gravity were obtained in the previous formalism defining
$\omega=-\phi/2$. 
We shall impose that, in order to recover the Einstein's regime
at large times and to be consistent with solar system tests of gravity,
$\dot\frac{(G^{-1})^4}{(G^{-1})^3} \rightarrow 0$ 
when $t \rightarrow \infty$
\cite {Ritis2}. This condition is satisfied when $G^{-1}$ 
tends to a constant without
asymptotic variations in the first derivative. 
This {\it matching approximation} carries the factor $\Gamma$ in the previous
equations to $\Gamma_a$, given by:

\begin{equation}
\Gamma_a=\left[ \frac 1\phi \frac{d\omega }{d\phi }-\frac
\omega {\phi ^2}-\frac 1G\frac{dG}{d\phi }\frac \omega \phi \right]
\end{equation}
Finally, the NMC slow roll equations may be written as 

\begin{equation}
\label{SR1-NMC}
3H\dot\sigma = - V^{\prime}(\sigma)
\end{equation}

\begin{equation}
\label{SR2-NMC}
3H^2=-\frac 14 \dot\phi ^2 G
+8\pi  G V(\sigma)
\end{equation}

\begin{equation}
\label{SR3-NMC}
\dot \phi ^2 \frac 1G \frac{dG}{d\phi}+
 3 H\dot\phi \left[ -1 +\frac
3{G^3}\left( \frac{dG}{d\phi }\right) ^2\right]=
-\frac 1G\frac{dG}{d\phi }32\pi V(\sigma)
\end{equation}

A further application of the slow roll condition on (\ref{SR3-NMC}) allows
to write it as

\begin{equation}
\label{SR3'-NMC}
3 H\dot\phi \left[ -1 +\frac
3{G^3}\left( \frac{dG}{d\phi }\right) ^2\right]=
-\frac 1G\frac{dG}{d\phi }32\pi V(\sigma)
\end{equation}
Using (\ref{SR2-NMC}) in (\ref{SR3'-NMC}) and suppossing that $\phi^2 G$ does not
diverge we obtain an additional product of the slow roll equations:

\begin{equation}
\label{SR3'''-NMC}
3 H\dot\phi \left[ -1 +\frac
3{G^3}\left( \frac{dG}{d\phi }\right) ^2\right]=
-\frac 1G\frac{dG}{d\phi }12 H^2 G^{-1}
\end{equation}
Note that the second term on the left behaves as $-9 \frac 1G \frac{dG}{d\phi}
H \dot{\left(G^{-1}\right)}$ while that on the right does as 
$-12 \frac 1G \frac{dG}{d\phi} H^2 G^{-1}$ and so we are allowed to have another
simplification

\begin{equation}
\label{SR3''-NMC}
3 H\dot\phi =
\frac{1}{G^2}\frac{dG}{d\phi}12 H^2
\end{equation}
This last equation may be readily integrated as 

\begin{equation}
\label{fi de a}
\int d\phi \frac {-1}{\frac{dG^{-1}}{d\phi}}=4 \ln \left( \frac{a}{a_0 }\right)
\end{equation}
with $a_0$ a constant of integration; thus giving $\phi=\phi(a)$
for every selection which makes invertible the result of the integral and 
$a=a(\phi)$ always.

To find $\phi=\phi(t)$ it is necessary to specify the form of the potential.
Let us consider some common cases.

\subsection{$V(\sigma)=V_0=constant$}

The case of constant potential allows complete integration. From the previous 
paragraph we have

\begin{equation}
3H^2G^{-1}=8\pi V_0
\end{equation}
Replacing the value of $H$ in (\ref{SR3''-NMC}) we obtain

\begin{equation}
\label{fi de t}
\int d\phi \frac {-1}{G^{1/2}\frac{dG^{-1}}{d\phi}}=
4 \left( \frac{8\pi V_0}{3}\right)^{\frac12} (t-t_0)
\end{equation}
with $t_0$ a constant of integration. Defining the gravitational theory
by giving the form of $G$, (\ref{fi de t}) gives $t=t(\phi)$ and, in the cases
in which inversion is possible, $\phi=\phi(t)$. In such cases, it will be 
also possible then, to replace in (\ref{fi de a}) to have $a=a(t)$.

We see that the procedure followed here continues the line of that presented 
by Barrow \cite{BarrowSR} in the case of given coupling of a BD theory; 
and as that,
this enable to see the form of inflation (if any) in a particular gravity theory
by integrating and inverting two differential equations. 

\subsection{$V(\sigma)=V_0 \exp{(-\lambda\sigma)}$; $V_0,\lambda=constants$}

With this exponential form for the inflaton potential the slow roll equations become

\begin{equation}
\label{SR1-exp}
3H\dot\sigma =  \lambda V_0 \exp{(\lambda\sigma)}
\end{equation}

\begin{equation}
\label{SR2-exp}
3H^2 G^{-1}=8\pi V_0 \exp{(-\lambda\sigma)}
\end{equation}

\begin{equation}
\label{SR3-exp}
3 H\dot\phi \left[ -1 +\frac{3}{G^3}
\left( \frac{dG}{d\phi }\right) ^2\right]=
-\frac 1G\frac{dG}{d\phi }32\pi V_0 \exp{(\lambda\sigma)}
\end{equation}
Defining as in \cite{BarrowSR} a new time coordinate as

\begin{equation}
t=\int 3H d\eta
\end{equation}
the equations are integrable again:

\begin{equation}
\sigma(\eta)= \frac1\lambda \ln \left[ \lambda^2 V_0 (\eta +\eta_0) \right] 
\end{equation}

\begin{equation}
\label{ln-exp}
\int d\phi \frac {-1+\frac{3}{G^3}\left(\frac{dG}{d\phi}\right)^2}
{G\frac{dG^{-1}}{d\phi}}=
32\pi \lambda^{-2} \ln \left[\eta +\eta_0 \right] 
\end{equation}
Defing $G$ and inverting (\ref{ln-exp}) to get $\phi$ as a function of $\eta$
one can use the slow roll equation (\ref{SR2-exp}) 
to finally obtain the behavior of $H$.

\subsection{$V(\sigma)=V_0 \sigma^{2r}$; $V_0,r=constant$ and $r \neq 1$}

With the same time variable, we have now:
     
\begin{equation}
\label{SR1-pot}
\sigma^\prime = - 2r V_0 \sigma^{2r-1}
\end{equation}

\begin{equation}
\label{SR2-pot}
3H^2 G^{-1}=8\pi V_0 \sigma^{2r}
\end{equation}

\begin{equation}
\label{SR3-pot}
\phi^\prime \left[ -1 +\frac
3{G^3}\left( \frac{dG}{d\phi }\right) ^2\right]=
-\frac 1G\frac{dG}{d\phi }32\pi V_0 \sigma^{2r}
\end{equation}
and the solutions are:

\begin{equation}
\sigma(\eta)=\left[ 4r (1-r) V_0  (\eta_0 -\eta)\right]^{\frac{1}{2(1-r)}} 
\end{equation}

\begin{equation}
\label{int-pot}
\int d\phi \frac {-1+\frac{3}{G^3}\left(\frac{dG}{d\phi}\right)^2}
{G\frac{dG^{-1}}{d\phi}}=
32\pi V_0 \left[4r (1-r) V_0 \right]^{\frac{r}{1-r}}
(\eta_0 -\eta)^{\frac{1}{(1-r)}}
\end{equation}
with $\eta_0$ a constant of integration.

\section{Slow Roll Solutions: Examples}

Let us first show how the formalism works
for integrable and invertible examples
which do not require numerical recipes and admit comparison
with previous works.

\subsection{$G^{-1}=\phi^n$, $n\neq 2,4$  and  $V=V_0=constant$}

We are dealing with the lagrangian density given by

\begin{equation}
L=\phi^n R+\frac12  \phi_{,\mu} \phi^{,\mu}+16 \pi L_m 
\end{equation}
Equation (\ref{fi de a}) may be integrated to give

\begin{equation}
\label{35}
\phi(a)^{-n +2}=-4n (2-n) \ln (a/a_0)
\end{equation}
Equation (\ref{fi de t}) may also be integrated: 

\begin{equation}
\phi(t)^{-n/2 +2}=-4n \left(-\frac{n}{2}+2\right) \left( \frac{8 \pi V_0}{3}
\right)^{\frac12} t
\end{equation}
and so

\begin{equation}
\phi(t)\propto t^{\frac{2}{-n+4}}
\end{equation}
This last relationship, used in (\ref{35}) gives

\begin{equation}
\left(\frac{a(t)}{a_0} \right) \propto \exp
{\left[t^{2\left(\frac{-n+2}{-n+4}\right)}\right]}
\end{equation}
When $n$ grows, the behavior of the scale factor tends to 
$\left(\frac{a(t)}{a_0} \right) \propto \exp{\left[t^2\right]}$.
We have to consider now if these solutions are consistent with the slow roll
approximations. We see that $\frac{\dot \phi}{H\phi}\ll 1$ for all $n$
greater that 4 and lower than 2 and that the approximation 
fails in the interval $\left(2,4 \right)$. The slow roll solutions found
have the form of those of intermediate inflation, in which the scale 
factor expands more slowly than for De Sitter case but faster
than for power law inflation \cite{Solutions}. The consistency 
of the $\Gamma$ to $\Gamma_a$
passage may be also tested, we find that $\frac{dG^{-1}}{d\phi} 
\propto t^{-2 \frac{2-n}{4-n}}$ and so, it is a decreasing quantity
with $t$.

\subsection{$G^{-1}=\phi^2$, $V=V_0=constant$}

The lagrangian density given by

\begin{equation}
L=\phi^2 R+\frac12  \phi_{,\mu} \phi^{,\mu}+16 \pi L_m 
\end{equation}
was mainly studied by the Naples group in relation with scalar potentials
and inflation \cite{Ritis}. Using equation (\ref{fi de a}) we get

\begin{equation}
\phi(a)\propto  \left(\frac{a}{a_0}\right)^{-8}
\end{equation}
and with equation (\ref{fi de t})

\begin{equation}
\phi(t)=ct
\end{equation}
with $c$ a constant. It may be verified with these solutions 
that both, the universe is not in expansion and the slow roll 
condition is not satisfied; and so, they must be discarded.

\subsection{$G^{-1}=\phi^4$, $V=V_0=constant$}

Equation (\ref{fi de a}) gives in this case

\begin{equation}
\frac{a}{a_0} \propto \exp[ \phi^{-2} ]
\end{equation}
while (\ref{fi de t}),

\begin{equation}
\phi \propto \exp[ -t ]
\end{equation}
Thus, the behavior of the scale factor with time is given by

\begin{equation}
a \propto \exp \left[ \exp[ 2t ]\right]
\end{equation}
Here, the scale factor appears to exhibit an extreme form of inflation.
It may be proved that $|\frac{\dot\phi}{H\phi}| \ll 1$ and 
$\frac{d^2G^{-1}}{d\phi^2} \rightarrow 0$ when $t$ is large enough.
The evolution of $\sigma(\eta)$ can be found from equation (\ref{4})
in each of the cases previously analized since $\dot\sigma \propto a^{-3}$.

\subsection{Spaning of Intermediate Inflation Solutions}

We are willing now to study in a completely general form
which kind of couplings may allow intermediate inflation
behavior. For simplicity we shall use the case of constant 
potential. In our study, we shall have two important equations:
(\ref{fi de a}), which is potential independent, and 
(\ref{fi de t}). 
From the integration of (\ref{fi de a}),
it may inmediately be seen that, in order to have a
scale factor evolving as

\begin{equation}
a(t) \propto \exp{[t^m]}
\end{equation}
it is necessary to have $t^m= \int d\phi \frac{-1}{\frac{dG^{-1}}{d\phi}}$,
that is:

\begin{equation}
\left[ \int d\phi \frac {-1}{G^{1/2}\frac{dG^{-1}}{d\phi}} \right]^m=
\int d\phi \frac{-1}{\frac{dG^{-1}}{d\phi}}
\end{equation}
Deriving with respect to $\phi$ we get:

\begin{equation}
\label{vinculum}
m \int d\phi \frac {G^2}{\frac{dG}{d\phi}} =
G^{\frac12}\int d\phi \frac{G^2}{\frac{dG}{d\phi} G^{\frac12}}
\end{equation}
which is an integral vinculum over the functional form of $G(\phi)$.
Although highly non-linear, equation (\ref{vinculum}) is very
suggestive. It may be proved that $G(\phi)=\phi^\alpha$ with 
$m=\frac{2(2+\alpha)}{4+\alpha}$
is one of its 
solutions, as expected, since we find intermediate inflation behavior
for it in the previous section. This case was already studied
concerning the fullfilment of the slow
roll conditions. Also the choice $G(\phi)=\alpha\exp{[ \alpha\phi]}$, 
with $\alpha$
a constant, is a solution of (\ref{vinculum}). For such choice, using the
previously derived method we obtain:

\begin{equation}
\label{t cuadrado}
a\propto exp{[t^2]}
\end{equation}
and

\begin{equation}
\phi\propto \frac{-2}{\alpha} \ln{[t]}
\end{equation}
It is woth recalling that the behavior of the sclae factor
in (\ref{t cuadrado})  coincides with that of $G=\phi^\alpha$
if $\alpha$ is large enough. The exponential form of $G$ is such that
the slow roll and the $\Gamma\rightarrow\Gamma_a$ conditions
are fulfilled.

\section{Discussion}

Starting from the hyperextended approach of scalar-tensor theories,
we have formulated the slow roll approximation in non-minimally
coupled gravity. The solutions were given in the form of two integrals 
that may be solved analytically in some cases and numerically in all cases.
These integrals give the behavior of $\phi$ as function of $a$ and $t$.
Properly inversion (again analytically or numerically) gives so, 
$a$ and $\phi$
as functions of time. We studied constant, power law and 
exponential
forms for the inflaton potential. Some examples that show intermediate 
inflation were presented for power law non-minimally couplings
in agreement with the results of power-law couplings in BD theories.
An integral vinculum for the functional form of $G$ was stablished in order to develop
intermediate inflation behavior and some examples were provided.

\section{Acknowledgments}

This work was partially supported by CONICET.
Valuable conversations with H. Vucetich are acknoledged.


\begin{thebibliography}{99}

\bibitem{Torres} D.F. Torres and H. Vucetich. At press in Phys. Rev. D.
See also N. Sakai and K. Maeda, Prog. of Theor. Phys. {\bf 90}, 1001, 1993

\bibitem{LiddleWands}  A.R. Liddle and D. Wands, Phys. Rev. D {\bf 45}, 2665
(1992)


\bibitem{Solutions}  D. La and P.J. Steinhardt, Phys. Rev. Lett. {\bf 62}, 376
(1989);  P.J. Steinhardt and F.S. Ascetta, Phys. Rev. Lett. {\bf %
64}, 2470 (1990)
J.D. Barrow and J.P. Mimoso, Phys. Rev. D {\bf 50}, 3746
(1994);
J.D. Barrow, Phys. Lett. B {\bf 235}, 40 (1990);                    
J.D. Barrow and P. Saich, {\it ibid}. {\bf 249}, 406 (1990)
J. Garcia Bellido and M. Quir\'os, Phys. Lett. B {\bf 243}, 45 (1990) 

\bibitem{SR} See for example: P. Coles and F. Lucchin, {\it Cosmology,
John Wiley 1995}. For a more rigurous approach: A. Liddle, P. Parsons
and J.D. Barrow, Phys. Rev. D. {\bf 50}, 7222 (1994)

\bibitem{Garcia} J. Garc\'ia-Bellido, A. Linde and D. Linde, Phys. Rev. D
{\bf 50}, 730 (1994)                    

\bibitem{BarrowSR} J.D. Barrow, Phys. Rev. D {\bf 51}, 2729 (1994)                    

\bibitem{Ritis2}  S. Capozziello, R. de Ritis and P. Scudellaro, Phys. Lett. 
A {\bf 188}, 130 (1994). 



\bibitem{Ritis}  S. Capozziello and R. de Ritis, Phys. Lett. A {\bf 195}, 48
(1994). See also: {\it ibid}. {\bf 177}, 1 (1993) and references therein.

\end{thebibliography}
\end{document}